\documentclass[a4paper,fleqn]{cas-dc}

\usepackage[authoryear,longnamesfirst]{natbib}
\usepackage[utf8]{inputenc}
\usepackage[outdir=./]{epstopdf}

\def\tsc#1{\csdef{#1}{\textsc{\lowercase{#1}}\xspace}}
\tsc{WGM}
\tsc{QE}

\begin{document}
\let\WriteBookmarks\relax
\def\floatpagepagefraction{1}
\def\textpagefraction{.001}

\shorttitle{yupi: Python library to handle trajectories}    

\shortauthors{A. Reyes \emph{et. al}}  

\title [mode = title]{yupi: Generation, Tracking and Analysis of Trajectory data in Python}  





\author[1]{A. Reyes}[auid=000, orcid=0000-0001-7305-4710]





\credit{Design of methods for generating, analyzing and visualization of trajectories}

\affiliation[1]{organization={Group of Complex Systems and Statistical Physics, University of Havana},
            addressline={San Lázaro esq. L, Vedado}, 
            city={La Habana},
            postcode={10400}, 
            country={Cuba}}

\author[2]{G. Viera-López}[auid=001, orcid=0000-0002-9661-5709]

\ead{gustavo.vieralopez@gssi.it}

\credit{Design of methods for tracking trajectories from video sources,
Conceptualization of the work}

\affiliation[2]{organization={Department of Computer Science, Gran Sasso Science Institute},
            addressline={Viale Francesco Crispi, 7}, 
            city={L'Aquila},
            postcode={67100}, 
            country={Italy}}

\author[1]{J.J. Morgado-Vega}[auid=000, orcid=0000-0001-6067-9172]





\credit{Main Software developer and maintainer}

\author[1]{E. Altshuler}[auid=003, orcid=0000-0003-4192-5635]





\credit{Methodology, Design of the examples}

\cortext[2]{Corresponding author}



\begin{abstract}
    The study of trajectories is often a core task in several research fields. 
    In environmental modelling, trajectories are crucial to study 
    fluid pollution, animal migrations, oil slick patterns or land 
    movements. In this contribution, we address the lack of standardization 
    and integration existing in current approaches to handle trajectory data. 
    Within this scenario, challenges extend from the extraction of a trajectory 
    from raw sensor data to the application
    of mathematical tools for modeling or making inferences about 
    populations and their environments. This work introduces a generic framework 
    that addresses the problem as a whole, i.e., a software library to handle 
    trajectory data. It contains a robust tracking module aiming at making data 
    acquisition handy, artificial generation of trajectories powered by 
    different stochastic models to aid comparisons among experimental 
    and theoretical data, a statistical kit for analyzing patterns in groups of 
    trajectories and other resources to speed up pre-processing of trajectory 
    data. 
    It is worth emphasizing that this library does not make 
    assumptions about the nature of trajectories (e.g., those from GPS), 
    which facilitates its usage across different disciplines. 
    We validate the software by reproducing key results when modelling   
    dynamical systems related to environmental modelling applications. An example 
    script to facilitate reproduction is presented for each case.
\end{abstract}





\ExplSyntaxOn
\keys_set:nn { stm / mktitle } { nologo }
\ExplSyntaxOff

\maketitle

\section*{Software Availability}\label{sec:availability}

\noindent
\textbf{Software name:} yupi \\
\textbf{Developers:} A. Reyes, G. Viera-López, J.J. Morgado\\
\textbf{First release:} 2021\\
\textbf{Program language:} Python\\
\textbf{License:} MIT\\
\textbf{Available at:} \\
\url{https://github.com/yupidevs/yupi} \\
\url{https://pypi.org/project/yupi/} \\
\textbf{Documentation:} \\
\url{https://yupi.readthedocs.io/en/latest/} \\
\textbf{Examples:} \\
\url{https://github.com/yupidevs/yupi_examples}

\section[Introduction]{Introduction} \label{sec:intro}

Environmental modelling, as many other fields of science, has been vastly 
impacted by a huge availability of mobile tracking sensors. The subsequent increase of
accessible trajectory data has lead to an uprising demand of trajectory analysis techniques.
For example, in Community Ecology and Movement Ecology different trajectory-based 
research is well developed \citep{de2019trajectory,demvsar2015analysis}.
Likewise, Group-Based Trajectory 
Modeling (GBTM), a statistical methodology for analyzing developmental trajectories, 
has been used in the study of restored wetlands \citep{matthews2015group}. Moreover, 
trajectory analysis has impacted the integration of land use and land cover data 
\citep{zioti2022platform} as well as oil spill environmental models for predicting oil 
slick trajectory patterns \citep{balogun2021oil} and pollution transients models
\citep{okamoto1987trajectory}. Furthermore, in the context of animal behavior, 
appropriate handling of trajectory data has allowed the characterization of behavioral 
patterns within a vast sample of organisms, ranging from microorganisms and cells 
\citep{figueroa2020coli,altshuler2013flow} to insects with a large impact in the 
environment, such as leaf-cutter ants
\citep{hu2016entangled,tejera2016uninformed}. This overwhelming increase on 
trajectory-related applications suggests to explore the available 
frameworks devoted to handle trajectory data.


Trajectory analysis software have been designed to address problems
in specific research fields (e.g., molecular dynamics \citep{roe2013ptraj,
kruger1991simlys}; modelling, transformation and visualization of urban
trajectory data \citep{shamal2019open}; animal trajectory analysis
\citep{mclean2018trajr} and human mobility analysis \citep{pappalardo2019scikit}).
For handling geo-positional trajectory data, a variety of tools has
been offered by different \textit{Python} libraries such as
\emph{MovingPandas} \citep{graser2019movingpandas}, \emph{PyMove}
\citep{sanches2019arquitetura, oliveira2019arquitetura} and \emph{Tracktable}
\citep{tracktable}. More recently, \emph{Traja} \citep{shenktraja} provided a
more abstract tool set for handling generic two-dimensional trajectories,
despite being focused around animal trajectory analysis. In the field of
Astrodynamics  high-level software has been provided by \textit{Julia}.
\emph{SatelliteToolbox.jl} is perhaps the most comprehensive astrodynamics
package available in \textit{Julia}, which is provided alongside the
in-development trajectory design toolkit, \emph{Astrodynamics.jl}
\citep{astrodynamics2013frazer}. In this regard, a programming toolkit
specialized in the generation, optimization, and analysis of orbital
trajectories has been published as \emph{OrbitalTrajectories.jl}
\citep{padilha2021modern}. \textit{R} language has been widely exploited as
well. For an excellent review and description of \textit{R} packages for
movement, broken down into three stages: pre{\textendash}processing,
post{\textendash}processing and analysis, see \citep{joo2020navigating}.

As a consequence of the specificity of existing frameworks, there is a 
wide diversity of software to address specific trajectory-related tasks, 
but a standard library for handling trajectories in an abstract manner
isn't available yet. For instance, most of existing software only address
two-dimensional trajectories or trajectories limited to a fixed number 
of dimensions. Moreover, they typically rely on different data structures
to represent a trajectory. 

In order to tackle these limitations, in this work we offer \emph{yupi}, a
general purpose software for handling trajectories regardless their nature. Our 
library aims to provide maximum abstraction from problem-specific details by
representing data in a compact and scalable manner and automating typical
tasks related to trajectory processing. At the same time, we want to encourage
the synergy among already available software. For this purpose, we also
provide tools to convert the trajectory objects used in our library into the
data structures used by other available frameworks, and vice-versa.

The software is the result of the experience gathered by the research our group has systematically conducted in the past 
few years regarding analysis and modelling of complex systems and visual 
tracking techniques in laboratory experiments. We believe that the field of environmental 
modelling is a strong candidate to showcase our library due to the wide 
variety of trajectory-related problems from different natures.

The manuscript is presented as follows: In Section \ref{sec:methods}, we 
describe the structure of the library, review basic concepts 
regarding trajectories and present the way \emph{yupi} handles them. Section 
\ref{sec:examples} presents applications that use trajectory analysis in diverse environmental modelling scenarios.
Finally, 
we summarize the work emphasizing the main contributions of \emph{yupi} and 
highlighting its current limitations.

\section{Software}\label{sec:methods}

Since \emph{yupi} aims to become a standard library to handle a wide spectrum of
tasks related to trajectories, all the components of the library share the
usage of a unified representation of \textbf{Trajectory} objects as the standard
structure to describe a path. Then, task-specific modules were conceived to
boost the processes of gathering, handling and analyzing trajectories. 

\begin{figure}[t]
   \centering
   \includegraphics[width=1\linewidth]{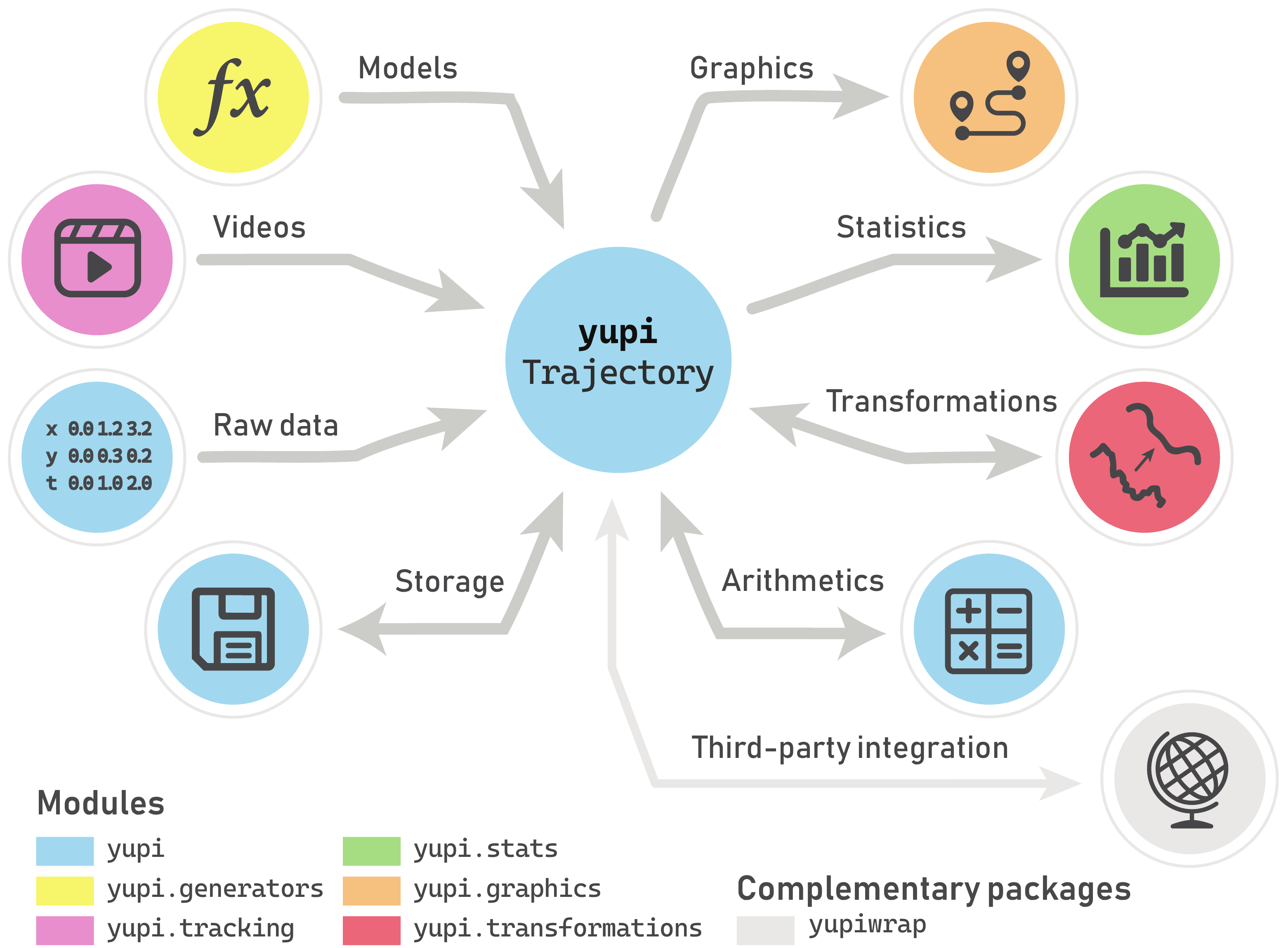}
   \caption{Visual representation of the internal structure of \emph{yupi}. Each
   possible path in the graph represents a possible application of \emph{yupi}.
Node colors resemble the internal \emph{yupi} module containing the specific
tool, with exception of gray, that represents the complementary software
package \emph{yupiwrap}, designed to ease the conversion process among
\emph{yupi} and third-party software.}
   \label{fig:A}
\end{figure}

The core module,
\textbf{yupi}, hosts the \textbf{Trajectory} class. It includes required
resources for arithmetic operations among trajectories and its storage on disk.
The library has six basic modules operating on \textbf{Trajectory} objects (see
Figure \ref{fig:A}). Artificial (i.e., simulated) trajectories with custom
mathematical properties can be created with \textbf{yupi.generators}. Data can be
extracted from videos using the \textbf{yupi.tracking} module. Regardless the
origin of a given trajectory, it can be altered using the
\textbf{yupi.transformations} module. Tools included in \textbf{yupi.stats} allow
the statistical analysis of an ensemble of trajectories. The module
\textbf{yupi.graphics} contains visualization functions for trajectories and its
estimated statistical quantities. In addition to \emph{yupi} internal modules,
we provide a complementary software package named \emph{yupiwrap} designed
exclusively to enable data conversion among \emph{yupi} and third-party
software. Next, we present each module of \emph{yupi} providing a brief description 
of its functionalities.

\subsection{Core module}\label{sub:core}
Empirically, a trajectory is the path that a body describes through space. More
formally, it is a function $\mathbf r(t)$, where $\mathbf r$ denotes position
and $t$, time. Here, $\mathbf r$ extends from the origin of an arbitrary
reference frame, to the moving body. Consequently, the core module
contains the 
class \textbf{Trajectory} to represent a moving object described by
some position 
vector, $\mathbf r(t)$, of an arbitrary number of dimensions. 

Since time is a continuous, machines have to deal with a discretized (i.e.,
sampled) version of the trajectory, $\mathbf r(t)$. As soon as one considers a
sampled trajectory, it always requires an associated time vector,
$\mathbf t=(t_1,...,t_n)^\intercal$, where each $t_i$ represents the
timestamp of the $i$-th sample and $n$ is the total number of samples. For
brevity, the sampled trajectory is often referred to as trajectory as well, so
we may use either term. For instance, in the typical 3-dimensional case, a 
trajectory can be defined by the vector $\mathbf r_i = (x_i, y_i, z_i)^\intercal$, 
where each component denotes a spatial coordinate.

The core module defines the way to retrieve specific quantities from a trajectory 
such as position components or velocity time series. It also defines operations 
among trajectories such as addition, scaling or rotation. In addition, storage 
functionalities are provided for different importing/exporting formats. Resources 
from this module can be imported directly from \emph{yupi} and will be summarized next.




\subsubsection{Vector objects}\label{sub:vector_objs}

It is very common to refer to position or velocity as a vector that changes 
through time. According to this, a \textbf{Vector} class was created to store 
all the time-evolving data in a trajectory. Iterating over each sample of these 
\textbf{Vector} time series one can get the vector components at specific 
time instants.

This class was implemented by wrapping the \emph{numpy} \textbf{ndarray} type. The
main reason that motivated this choice, along with all the benefits from the
\textbf{ndarray} class itself, was to gain verbosity over the usual operations on
a vector. For instance, getting a specific component of a vector, the
differences between its elements, or even calculating its norm, can be done
with a vector instance by accessing properties such as: \textbf{norm} or
\textbf{delta}. In addition, properties \textbf{x}, \textbf{y}, \textbf{z} 
allow acquiring data from one specific axis in multidimensional vectors. 

Although users may not directly instantiate \textbf{Vector} objects, these are
used all along the library to represent every time-evolving data one could get 
from a trajectory such as position, velocity, acceleration and time itself.

\subsubsection{Trajectory objects}\label{sub:trajectory_objs}

A \textbf{Trajectory} object is \emph{yupi}'s essential structure. Its time evolving 
data, stored as \textbf{Vector} objects, can be accessed through
the attributes \textbf{t}, \textbf{r}, \textbf{v} and \textbf{a}, standing 
for time, position, velocity and acceleration, respectively.

Trajectory data is typically stored in different manners, e.g., a single sequence of 
$d$-dimensional points where each point represents the position at a time instant 
or, alternatively, $d$ sequences of position components. Regardless of 
the input manner, \emph{yupi} offers the way to create \textbf{Trajectory} 
objects from raw data.

By default, trajectories will be assumed to be uniformly spaced every 1 unit of 
time. However, custom time information can also be 
supplied. Then, position and time are used to automatically estimate velocity 
and acceleration according to one of the supported numerical methods: linear 
finite differences and the method proposed by \citep{fornberg1988generation}.

\textbf{Trajectory} objects can be shifted or scaled by performing arithmetic 
operations on either all position components of a subset of them. For the 
specific case of 2- or 3-dimensional trajectories, rotation methods were 
conveniently implemented to ease visualization tasks, named \textbf{rotate\_2d} 
and \textbf{rotate\_3d}.

Furthermore, operations among trajectories are also defined.
Trajectories with the same dimension and time vector can be added, subtracted or multiplied together. These operations are defined point-wise and can be used via the conventional operators for 
addition, subtraction and multiplication: \textbf{+}, \textbf{-} and \textbf{*}.


\subsection{Generators module}\label{sub:generating}

The usage of randomly generated data is common in different research approaches
related to trajectory analysis \citep{tuckerman2010statistical}.
In this section we tackle three classical models
that usually explain (or serve as a framework to explain) a wide number of 
phenomena connected to Biology, Engineering and Physics: Random Walks \citep{pearson1905problem}, the Langevin model \citep{langevin1908theory} and Diffusing-Diffusivity model \citep{chechkin2017brownian}. 

In \emph{yupi}, the aforementioned models are implemented by inheritance of an
abstract \textbf{Generator} class. 
Any \textbf{Generator} object must be used specifying four parameters that characterize numerical properties 
of the generated trajectories: \textbf{T} (total time), \textbf{dim} (trajectories dimension), \textbf{N} (number of trajectories to be generated) and
\textbf{dt} (time step). Additionally, a \textbf{seed} parameter can be specified 
to initialize a random number generator (\textbf{rng}) which is used locally 
to reproduce the same results without changing the global seed.

Next, we will briefly describe the foundations of each model implemented
in \emph{yupi} and explain how to use them to generate ensembles of trajectories like the ones sketched in Figure \ref{fig:B}a.

\subsubsection{Random walk}\label{subsub:randomwalk}

A Random Walk is a random process that, in a $d$-dimensional space ($d \!\in\!
\mathbb{N}^+$), describes a path consisting in a succession of independent
random displacements. Since the extension to more than one dimension is
straightforward, we shall formulate the process in its simplest way:

Let $Z_1, ..., Z_i, ..., Z_{n}$ be independent and identically distributed
random variables (r.v.'s) with $P(Z_i=-1)=q$, $P(Z_i=0)=w$ and $P(Z_i=1)=p$,
and let also $X_i = \sum_{j=1}^{i} Z_j$. Interpreting $Z_i$ as the displacement at the
$i$-th time instant, the collection of random positions $\{X_i, 1 \le i \le n
\}$ defines the well known random walk in one dimension.

It is possible to extend this definition by allowing the walker to perform
displacements of unequal lengths. Hence, if we denote by $L_i$ the variable that
accounts for the length of the step at the $i$-th time instant, the position
will be given by:
\begin{equation}
   X_i = \sum_{j=1}^{i} L_j Z_j, \qquad i = 1,2,...,n
  \label{eq:def-random-walk}
\end{equation}

A process governed by Equation~\ref{eq:def-random-walk} in each axis is a
$d$-dimensional Random Walk. Note that our definition is slightly more general
than the classical, i.e., it allows the walker to remain at rest in a node of a
network that is not necessarily evenly spaced. These generalized versions are
often known as Lazy Random Walks \citep{lawler2010random} or Random Walks with
multiple step lengths \citep{boczkowski2018random}.

We define a trajectory by having a position vector whose components are
described by Equation~\ref{eq:def-random-walk}. For instance, in the 3-dimensional
case, $\mathbf r_i = (X_i^{(1)}, X_i^{(2)}, X_i^{(3)})$.

In \emph{yupi}, this model is accessible through the \textbf{RandomWalkGenerator} class. 
To use it, the probabilities $q$, $w$ and $p$ need to be defined for each dimension:
\begin{verbatim}
    prob = [[.5, .1, .4],   # x-axis
            [.5,  0, .5]]   # y-axis
\end{verbatim}

Then, trajectories are generated as:
\begin{verbatim}
from yupi.generators import RandomWalkGenerator
rw = RandomWalkGenerator(T, dim, N, dt, prob)
trajs = rw.generate()
\end{verbatim}

In this case, the variable \textbf{trajs} contains a list of \textbf{N} generated \textbf{Trajectory} objects. Note that the first 4 parameters passed to the \textbf{RandomWalkGenerator} are required by any kind of generator in \emph{yupi} as explained in the beginning of this section.

\subsubsection{Langevin model}\label{subsub:langevin}

An Ornstein-Uhlenbeck process \citep{uhlenbeck1930theory}, well known by its
multiple applications to describe processes from different fields 
\citep{lax2006random}, is defined in the absence of drift by the linear 
stochastic differential equation:
\begin{equation}
   dv = -\gamma \, v \,dt + \sigma \, dW
  \label{eq:oh-process}
\end{equation}

\noindent
where $\gamma$ and $\sigma$ are positive constants and $W$ is a Wiener process.
Equation~\ref{eq:oh-process} is also written as a Langevin equation
\citep{langevin1908theory}, which in the multi-dimensional case takes the form:
\begin{equation}
      \frac{d}{dt}\mathbf v(t) = -\gamma \mathbf v(t) + \sigma \boldsymbol \xi(t)
     \label{eq:langevin-eq}
\end{equation}

where $\boldsymbol \xi(t)$ is a white noise; $\sigma$, the scale noise
parameter; and $\gamma^{-1}$, a characteristic relaxation time of the process.

In trajectory analysis, $\mathbf v(t)$ is intended to denote velocity, so the
position vector is given by:
\begin{equation}
   \mathbf{r}(t) = \int_{0}^{t} \mathbf{v}(t') \,dt'
  \label{eq:r-from-v}
\end{equation}

Equations~\ref{eq:langevin-eq} and \ref{eq:r-from-v} can be solved
numerically if the initial conditions $\mathbf v(0) = \mathbf v_0$ and
$\mathbf r(0) = \mathbf r_0$ are known.


\textbf{LangevinGenerator} is the class offered by \emph{yupi} to
generate trajectories that follow this model. By conveniently setting the 
parameters described above (i.e., $\gamma$, $\sigma$, $\mathbf v_0$ and 
$\mathbf r_0$) several real-life scenarios can be modeled. In Section
\ref{sub:ejlysozyme}, a Langevin model is used to simulate 
the motion of a Lysozyme molecule in an aqueous medium.

\subsubsection{Diffusing-Diffusivity model}\label{subsub:diffdiff}
Slow environmental relaxation has been found in soft matter for colloidal 
particles diffusing in an environment of biopolymer filaments and 
phospholipid tube assemblies \citep{wang2012brownian}. Non-Gaussian 
distribution of increments was observed even when the diffusive dynamics 
exhibit linear growth of the mean square displacement. A model framework of 
a diffusion process with fluctuating diffusivity that reproduces this 
interesting finding has been presented as Diffusing Diffusivity Model. 
Namely:
\begin{subequations}
  \begin{align}
    \frac{d}{dt} \mathbf r(t) &= \sqrt{2D(t)} \,\boldsymbol\xi(t)
    \\
    D(t) &= \mathbf Y^2(t)
    \\
    \frac{d}{dt} \mathbf Y(t) &= -\frac{1}{\tau} \mathbf Y(t) + \sigma \boldsymbol \eta(t)
  \end{align}
 \label{eq:diffdiff}
\end{subequations}

where $\boldsymbol\xi(t)$ and $\boldsymbol\eta(t)$ are Gaussian white noises
and $D(t)$, the diffusion coefficient, is a random function of time expressed
as the square of the auxiliary variable, $\mathbf Y(t)$. 

In other words, the coupled set of stochastic differential equations
(\ref{eq:diffdiff}) predicts a Brownian but Non-Gaussian diffusion, where the
position, $\mathbf r(t)$, is described by an over-damped Langevin equation with
the diffusion coefficient being the square of an Ornstein-Uhlenbeck process. The
model has been discussed and solved analytically by
\citep{chechkin2017brownian,thapa2018bayesian}.

The class devoted to generate trajectories modeled by the Equations
\ref{eq:diffdiff} is called \textbf{DiffDiffGenerator}. Implementation details can be found in the software documentation.

\begin{figure*}[t]
   \centering
   \includegraphics[width=1\linewidth]{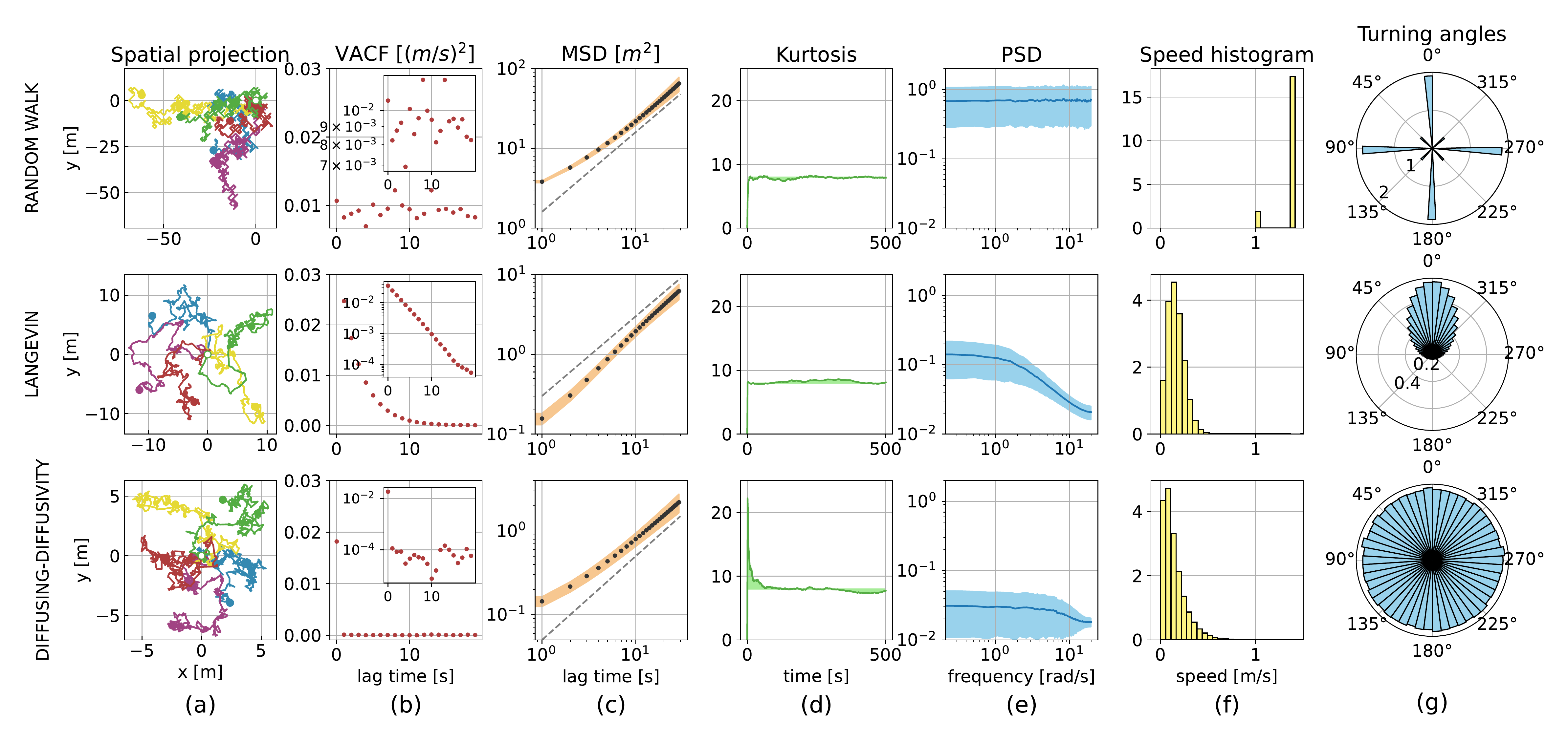}
   \caption{Statistical analysis of three generated ensembles of N=1000 two-dimensional trajectories. Rows listed from top to bottom correspond to an ensemble generated using a Random Walk, Langevin and Diffusing Diffusivity models, respectively. (a) Spacial projection of five trajectories. (b) Estimated Velocity autocorrelation function and (c) Mean Squared Displacement as a function of the lag time. (d) Kurtosis as a function of time. (e) Power Spectral Density. (f) Speed histograms. (g) Turning angle probability densities.}
   \label{fig:B}
\end{figure*}

\subsection{Stats module}\label{sub:soft_stats}

The library provides common techniques based on the 
mathematical methods that researchers frequently use when analyzing trajectories. 
This section presents an overview on how to compute observables to 
describe them. Velocity autocorrelation function, kurtosis or mean square displacement are some typical examples. 

Most of these observables were originally defined using ensemble averages (i.e.,
expected values at a given time instant). 
However, under the assumption of ergodicity\footnote{Ergodicity is the property of a 
process in which long-time
averages of sample functions of the process are equal to the corresponding 
statistical or ensemble averages.},
it is possible to compute the observables using time averages (i.e., by averaging 
the quantities among time intervals instead). In
\emph{yupi} the observables can be computed using both approaches. 
In the following subsections, we will denote by $E[\cdot]$ (i.e., expected value) 
the ensemble average and $\langle\cdot\rangle$ the time average.

\subsubsection{Velocity autocorrelation function}\label{subsub:VACF}


The velocity autocorrelation function (VACF) is defined as the 
ensemble average of the product of velocity vectors at any two instants of 
time. Under stationary conditions, the definition is typically relaxed to the one of 
Equation~\ref{eq:def-vacf-avg-ensemble}, in which one of the vectors is 
the initial velocity. On the other hand, Equation~\ref{eq:def-vacf-time-avg-time} 
presents the way VACF is computed by averaging over time under the assumption 
that both averages are equivalent (i.e., ergodic assumption).
\begin{subequations}
    \begin{align}
        C_v(t) &= 
        E[\mathbf{v}(0) \cdot \mathbf{v}(t)]
        \label{eq:def-vacf-avg-ensemble}
        \\
        \begin{split}
            C_v(\tau) &= 
            \langle \mathbf{v}(t) \cdot \mathbf{v}(t+\tau) \rangle 
            \\
            &= \frac{1}{T-\tau} \int_0^{T-\tau} \mathbf{v}(t) \cdot \mathbf v(t+\tau) dt
        \end{split}
        \label{eq:def-vacf-time-avg-time}
    \end{align}
\end{subequations}

\noindent
Here, $\tau$ is 
the lag time, a time window swept along the velocity samples and $T$ is the 
elapsed time, where $\tau \ll T$. 

It should be noted that in Equation~\ref{eq:def-vacf-time-avg-time} VACF is
defined to be computed on a single trajectory, unlike Equation
\ref{eq:def-vacf-avg-ensemble} that requires an ensemble. This
also applies to further statistical observables that can be computed using both
kinds of averaging procedures.

VACF quantifies the way in which the memory in the velocity decays as a
function of time \citep{balakrishnan2008elements}. Moreover, it can be
successfully used to analyze the nature of an anomalous diffusion process
\citep{metzler2014anomalous}.

From top to bottom, Figure \ref{fig:B}b shows VACF plots for Random Walk, Langevin and Diffusing Diffusivity generated ensembles. VACF scatter as an almost flat curve around zero for the top and bottom case, which indicates the memoryless nature of Random Walk and Diffusing Diffusivity processes. On the other hand, the center row shows an exponential decay, meaning that a Langevin model predicts some characteristic time that dominates the relaxation to equilibrium.

With \emph{yupi}, one can estimate the VACF of a collection of trajectories (e.g., the ensemble generated in Section \ref{subsub:randomwalk}) as:
\begin{verbatim}
    from yupi.stats import vacf
    trajs_vacf, trajs_vacf_std = vacf(
        trajs, time_avg=True, lag=25)
\end{verbatim}

\noindent
where \textbf{vacf} is the name of the function that computes the autocorrelation, 
the parameter \textbf{trajs} represents the array of trajectories and \textbf{time\_avg} 
indicates the method to compute the observable, i.e., averaging over time with a lag 
time defined by \textbf{lag}.

The computation of the remaining observables can be coded in a similar way. 
Next, for the sake of brevity, we will address only its theoretical foundations. 
In the software documentation, more examples can be found that make use of
all the observables.

\subsubsection{Mean square displacement}\label{subsub:MSD}

The mean square displacement (MSD) is defined in Equation
\ref{eq:def-msd-avg-ensemble} by an ensemble average of square displacements.
In addition, the time-averaged mean square displacement (TAMSD) is computed by a
moving average of the squared increments along a single trajectory. This is
performed by integrating over trajectory points separated by a lag time $\tau$
that is much smaller than the overall measurement time $T$ (Equation
\ref{eq:def-msd-avg-time}).
\begin{subequations}
    \begin{align}
        \delta^2(t) &= 
        E[{(\mathbf r(t) - \mathbf r(0))}^2]
        \label{eq:def-msd-avg-ensemble}
        \\
        \begin{split}
            \delta^2(\tau) &= 
            \langle {(\mathbf r(t+\tau) - \mathbf r(t))}^2 \rangle 
            \\
            &= \frac{1}{T-\tau} \int_0^{T-\tau} {(\mathbf{r}(t+\tau) - \mathbf r(t))}^2 dt
        \end{split}
        \label{eq:def-msd-avg-time}
    \end{align}
\end{subequations}

The MSD of a normal diffusive trajectory arises as a linear function of time.
Therefore, it is a typical indicator to classify processes far from normal 
diffusion. Moreover, MSD reveals what are the time scales that characterize 
different diffusive regimes.

In Figure \ref{fig:B}c a comparison of MSD plots is made for the three ensembles previously presented. Regardless of the model used, the same long time behavior can be perceived, i.e., the same scaling law arises for sufficiently long time scales. This can be seen while contrasting the MSD curves with the dashed line of slope equal to one, meaning that normal diffusion is achieved.

\subsubsection{Kurtosis}\label{subsub:kurtosis}

Another useful statistical observable is the kurtosis. Different formulations
for kurtosis have been proposed in the literature \citep{cain2017univariate}.
The common choice for the one-dimensional case is presented as the ensemble
average of Equation~\ref{eq:def-kurtosis-1d}, where $\mu$ stands for the mean
velocity and $\sigma$ the standard deviation. In addition, for the multivariate
case we use Mardia's measure \citep{mardia1970measures} as in
Equation~\ref{eq:def-kurtosis-Nd}. The vector $\boldsymbol{\mu}$ is the
$d$-dimensional mean velocity ($d>1$) and $\mathbf{\Sigma}$ the covariance
matrix. In both cases we have omitted the explicit dependence with time, but it
should be noted that the expected value is taken at a given instant. 
\begin{subequations}
   \begin{align}
      \kappa(t) &= E\left[\left( \frac{v - \mu}{\sigma} \right)^4\right]
     \label{eq:def-kurtosis-1d}
      \\
      \kappa(t) &= E[\{(\mathbf{v} - \boldsymbol{\mu})^\intercal \mathbf{\Sigma}^{-1} (\mathbf{v} - \boldsymbol{\mu})\}^2]
     \label{eq:def-kurtosis-Nd}
   \end{align}
  \label{eq:def-kurtosis}
\end{subequations}

The kurtosis measures the disparity of spatial scales of a dispersal process
\citep{mendez2016stochastic} and it is also an intuitive means to understand
normality \citep{cain2017univariate}.

Figure \ref{fig:B}d shows how kurtosis converges to a value close to $8$ regardless of the model used. This is a consequence of convergence to a Gaussian density and the fact that all three processes are two-dimensional. Moreover, just in the case of the Diffusing Diffusivity model a leptokurtic regime is observed, i.e., a regime in which $\kappa \gtrsim 8$. This means that some flat-tailed density (compared with the Gaussian) aroused first and $\kappa(t)$ provides a direct way to extract, apart from the crossover time, the correlation time of the diffusion coefficient.

\subsubsection{Power spectral density}\label{subsub:psd}

The Power Spectral Density, or Power Spectrum, (PSD) of a continuous-time
random process can be defined by virtue of the Wiener${-}$Khintchin theorem as
the Fourier transform $S(\omega)$ of its autocorrelation function $C(t)$:
\begin{equation}
   S (\omega) = \int_{-\infty}^\infty C(t) e^{-i \omega t} dt
\end{equation}

Power spectrum analysis indicates the frequency content of the process. 
The inspection of the PSD from a collection of trajectories enables
the characterization of the motion in terms of the frequency components. 

For instance, when analysing the ensembles represented in Figure \ref{fig:B}a, 
we notice important differences in their spectrum (see Figure \ref{fig:B}e). 
In the Langevin and Diffusing Diffusivity cases the PSD shows a decay for 
larger frequencies as opposite to the Random Walk model, in which all frequencies 
contribute equally, i.e., the spectrum is distributed uniformly.

\subsubsection{Histograms}\label{subsub:histograms}

Certain probability density functions can also be estimated from input
trajectories (e.g., velocity and turning angle distributions).

Speed probabilty density function is a useful observable to inspect jump length statistics. 
For instance, Figure \ref{fig:B}f reveals the discrete nature of the 
Random Walk and the rapidly decay of the tails for the Langevin and 
Diffusing Diffusivity plots, which is a typical indicator to discard anomalous 
diffusion models as candidate theories.

Figure \ref{fig:B}f shows turning angle distributions in polar axes. For the 
Random Walk model just few discrete orientations are available in contrast with the 
other two cases: a bell-shape around zero and a uniform distribution for the Langevin and Diffusing Diffusivity model, respectively.

\subsubsection{Other functionalities}\label{subsub:otherquantities}

In addition to the computation of statistical estimators, the \textbf{stats}
module of \emph{yupi} includes the \textbf{collect} function for querying specific
data from a set of trajectories. If one desires to obtain only position, velocity or
speed data from specific time instants, this function automatically iterates over the
ensemble and returns the requested data. Moreover, \textbf{collect} also gets
samples for a given time scale using sliding windows.

A more extensive showcase of this module can be seen in the examples provides as part of the Software Documentation.

\subsubsection{Graphics module}\label{sub:soft_graphics}

A set of pre-configured visualization functions are included as part of \emph{yupi}. Spatial projections can be visualized for the cases of 2- and 3-dimensional trajectories 
using \textbf{plot\_2d} and \textbf{plot\_3d} functions. For instance, each subplot in Figure \ref{fig:B}a is the outcome of \textbf{plot\_2d} for different ensembles.

In addition, specific plots were added to ease the visualization of the observables offered by the module \textbf{yupi.stats}). This customized plotting functions were 
designed to highlight statistical patterns following the commonly used standards in 
the literature (e.g., by default, plots of angle distributions are displayed in polar
coordinates and the y- and x-axis of the Power Spectral Density plots in logarithmic 
scale). All the plots in Figure \ref{fig:B}b-g were produced using the aforementioned functions.

All these functions were conceived to allow plots customization through case-specific parameters (e.g., PSD can be plotted as a function of the frequency or angular 
frequency). Moreover, since all the predefined plots were implemented over \emph{matplotlib}, 
the users can fully customize their plots via keyword arguments (\textbf{kwargs} parameter) 
that will override any default values imposed by the specific \emph{yupi} plotting
function.

\subsection{Transformation module}\label{sub:soft_transf}

The \emph{yupi.transformations} module can be used when the desired outcome is a  
``transformed'' version of a given trajectory that does not modify the trajectory 
itself. Since several methods of this kind can be applied from standard signal 
processing libraries (e.g., \emph{scipy.signal}) we kept this module simple. Therefore,  
we included mostly specific resources that were both useful in the context of trajectory analysis and uncommon in most popular signal processing libraries.

\subsubsection{Trajectory filters}\label{sub:filters}

The module \emph{scipy.signal} offers methods to convolve, to spline or to apply low-, band- 
and high-pass filters. However, we have included a convenient filter especially useful in 
the context of animal behavior, where the instant velocity vector is sometimes 
approximated to a local weighted average over past values \citep{li2011dicty}. This is 
presented as the convolution:
\begin{equation}
   \mathbf{v}_s(t) = \Omega \int_0^t e^{-\Omega(t - t')} \mathbf{v}(t') dt'
  \label{eq:exp-convolution}
\end{equation}

\noindent
where $\Omega$ is a parameter that accounts for the inverse of the time window over which 
the average is more significant. A filter defined by Equation \ref{eq:exp-convolution} 
preserves directional persistence and produces a smoothed version of a trajectory whose 
velocity as a function of time is given by $\mathbf v_s(t)$. Therefore, position can be 
recovered with the help of Equation~\ref{eq:r-from-v}.

This ``exponential-convolutional'' filter can be used as:
\begin{verbatim}
from yupi.transformations import (
    exp_convolutional_filter)
smooth_traj = exp_convolutional_filter(
    traj, ommega=5)
\end{verbatim}

Future releases of the library may include new filters required for specific applications in trajectory analysis. 

\subsubsection{Trajectory re-samplers}\label{subsub:samplers}

There are several applications that require trajectories to be sampled in specific time arrays.
The most obvious case is when a trajectory has a non-uniform time array and it is desired to
produce an equivalent trajectory sampled periodically on time. This can be achieved using the
\textbf{resample} function:
\begin{verbatim}
    from yupi.transformations import resample
    t1 = resample(traj, new_dt=0.3, order=2)
\end{verbatim}

Notice that, by default, the library uses a linear interpolation to resample the trajectory. 
However, the order of the estimation can be controlled using the \textbf{order} parameter. 

Equivalently, a new trajectory can be obtained for a given time array that is not required to
be uniformly sampled by specifying the time array itself as the $new\_t$ parameter instead of $new\_dt$ while
calling the \textbf{resample} function. 

We also included a simple sub-sampling method designed for uniformly-sampled trajectories. 
It produces trajectories that keep only a fraction of the original trajectory points. 
It can be used as:
\begin{verbatim}
    from yupi.transformations import subsample
    compact_traj = subsample(traj, step=5)
\end{verbatim}

\noindent
where the \textbf{step} parameter is specifying how many sample points will be skipped.

\subsection{Tracking module}\label{sub:tracking}

The tracking module contains all the tools related to retrieving trajectories 
from video inputs. Although \emph{yupi} works with trajectories of an arbitrary 
number of dimensions, the scope of this module is limited to two-dimensional 
trajectories due to the nature of video sources. However, inspired by several 
practical scenarios in which tracking techniques are required to extract 
meaningful information, we decided to include these tools as part of the library.

We will refer to tracking as the process of
retrieving the spatial coordinates of moving objects in a sequence of images.
Notice that this is not always possible for any image sequence. Some 
requirements should be met in order to extract meaningful information.

Along this section, we will assume that any video used for tracking purposes
was taken keeping a constant distance from the camera to the plane
in which the target objects are moving.

\subsubsection{Tracking of objects of interest}\label{subsub:objecttracking}

When following a given object in a
video, the aim of tracking techniques is to provide its position vector with
respect to the camera, which we will denote by $\mathbf{r}_i^\mathrm{(oc)}$
(superscript $(oc)$ stands for object-to-camera reference and subscript $i$ stands 
for the $i$-th frame the vector is referred to).

If the dimensions of the object being tracked are small enough (i.e.,
comparatively smaller than the distance covered in the whole trajectory) the
centroid of the object determines the only degrees of freedom of the movement
since orientation can be neglected. This point is frequently taken as the
position of the object. 

To determine the actual position of an object on every frame, a tracking
algorithm to segment the pixels belonging to the object from the background
is required. Five algorithms are implemented in \emph{yupi}: 
\textbf{ColorMatching}, \textbf{FrameDifferencing}, \textbf{BackgroundSubtraction}, 
\textbf{TemplateMatching} and \textbf{OpticalFlow}. A detailed explanation of
the basics of the aforementioned algorithms can by found in \cite{frayle2017chasing}.
All those algorithms attempt to
solve the same problem using very different strategies. Therefore, it happens often 
that under specific conditions one of them may outperform the others.

To speed up the tracking process, \emph{yupi} uses a region of interest (ROI) around the
last known position of the tracked objects. Then, on every frame, the algorithm
searches only inside this region instead of in the whole image. When it founds 
the new position, it updates the center of the region of interest for the next 
frame. When the tracking ends, the time evolution of the central position of
the ROI is used to reconstruct the trajectory of the tracked object. 

The library enables the extraction of \textbf{Trajectory} objects from videos
through \textbf{ObjectTracker} instances which are defined by a tracking algorithm 
and a ROI. For a given video, several objects can be tracked concurrently
using an independent \textbf{ObjectTracker} for each one of them. Finally, a
\textbf{TrackingScenario} is the structure that groups all trackers and iterates 
through the video frames while resolving the desired trajectories
for each tracker.

\subsubsection{Tracking a camera in motion}\label{subsub:cameratracking}

Sometimes the camera used to track the object under study is also in motion and
it is able to translate and to execute rotations around a vertical axis. Therefore,
knowledge regarding positions and orientations of the camera during the whole
experiment is required to enable a correct reconstruction of the trajectory
relative to a fixed coordinate system. One way to tackle this issue is
inferring the movement of the camera by means of the displacements of the
background in the image. Preserving the assumption that the camera only moves
in a plane parallel to the plane in which the objects are moving, the motion 
of the camera can be estimated by tracking a number of background
points between two different frames and solve the optimization problem of
finding the affine matrix that best transforms the set of points\footnote{An affine
transformation (or affinity) is the one that preserves collinearity and ratios
of distances (not necessarily lengths or angles).}. This means that for a
rotation angle $\theta$, a scale parameter $s$, and displacements $t_x$ and
$t_y$, an arbitrary vector $(x,y)$ will become 
\begin{equation}
   \begin{bmatrix}
      x' \\ y' \\ 1
   \end{bmatrix}
   =
   \begin{bmatrix}
      s \cos \theta & -\sin \theta & t_x \\
      \sin \theta & s \cos \theta  & t_y \\
      0 & 0 & 1
   \end{bmatrix}
   \begin{bmatrix}
      x \\ y \\ 1
   \end{bmatrix}
  \label{eq:affine_transformation}
\end{equation}

Therefore, the problem reduces to find the vector $(\theta, t_x, t_y, s)$ that
minimizes the least square error of the transformation. Under the assumption
that the camera is always at approximately the same distance from the
background, the scale parameter should be close to 1. As long as the mean
square error remains under a given threshold and the condition $s \approx 1$
holds, the validity of the estimation is guaranteed. Hence, the collection
$\{\theta_i, t_{xi}, t_{yi}\}$, where $i$ stands for the frame number, contains
all the information necessary to compute the positions and orientations of the
camera.

\subsubsection{Tracking objects and camera simultaneously}

As was mentioned above, the position of the object under study with respect to
the camera in the $i$-th frame has been denoted by
$\mathbf{r}_i^\mathrm{(oc)}$. If the camera also moves and we can follow
features on the background from one frame to the another, we are able to
calculate the parameters $(\theta_i, t_{xi}, t_{yi})$ of the affine matrix that
transforms the $(i{-}1)$-th into the $i$-th frame. 

The question we now face is:
How to compute the position $\mathbf{r}_i \equiv \mathbf{r}_i^\mathrm{(ol)}$ of
the object with respect to the frame of reference fixed to the lab in the
$i$-th frame?

Let us label as $\alpha_i$ the cumulative angular differences of the affine
matrix parameter $\theta$ until the $i$-th frame (Equation~\ref{eq:cum_theta}) and
$\mathbf{t}_i = (t_{xi}, t_{yi})^\intercal$  the vector of displacements of the
affine transformation. Let $\mathbf{R}$ be the rotation matrix (i.e., the
upper-left $2{\times}2$ block of the matrix in
Equation~\ref{eq:affine_transformation} when $s=1$). Then, the position of the
camera in the lab coordinate system, $\mathbf{r}_i^{(\mathrm{cl})}$, can be
determined in an iterative manner as follows:
\begin{subequations}
   \begin{align}
      \alpha_i &= \sum_{j=1}^{i} \theta_j
      \qquad
     \label{eq:cum_theta}
      \\
      \mathbf{r}_i^\mathrm{(cl)} &= \mathbf{R}^{-1}(\alpha_i) \cdot (-\mathbf{t}_i) + \mathbf{r}_{i-1}^\mathrm{(cl)}
     \label{eq:r-cl}
      \\
      \mathbf{r}_i &= \mathbf{R}^{-1}(\alpha_i) \cdot \mathbf{r}_i^\mathrm{(oc)} + \mathbf{r}_{i}^\mathrm{(cl)}
     \label{eq:r-ol}
   \end{align}
\end{subequations}

Therefore, the desired position of the object under study, $\mathbf{r}_i$, can
be computed by Equation~\ref{eq:r-ol} in terms of its position in the frame of
reference fixed to the camera, $\mathbf{r}_i^\mathrm{(oc)}$, the absolute
position of the camera, $\mathbf{r}_i^\mathrm{(cl)}$, and the camera
orientation, $\alpha_i$.

This whole process is simplified in \emph{yupi} by the \textbf{CameraTracker} class.
By default, \emph{yupi} assumes that the position of the camera remains fixed.
However, in order to estimate the motion of a camera and use it to retrieve the
correct position of the tracked objects, the user only needs to create a
\textbf{CameraTracker} object and pass it to the \textbf{TrackingScenario}.

\subsubsection{Removing distortion}

When recording video, the output images generally contain some distortion
caused by the camera optics. This is important when spatial measurements are
being done using videos or photographs. To correct these errors some
adjustments must be applied to each frame. Applying an undistorter function to
the videos is possible in \emph{yupi} using the \textbf{ClassicUndistorter} or the
\textbf{RemapUndistorted}\footnote{To instantiate one of the given undistorters a
camera calibration file is needed. This file can be created following the
steps explained in the documentation
\url{https://yupi.readthedocs.io/en/latest/api_reference/tracking/undistorters.html}.}.

\subsection{Integration with other software packages}\label{sub:soft_yupiwrap}

Along with \emph{yupi}, we offer a \textit{Python} library called \emph{yupiwrap} 
(available in \url{https://github.com/yupidevs/yupiwrap}), designed to ease the 
integration of \emph{yupi} with other libraries for handling trajectories. Two-way 
conversions between a given \emph{yupi} \textbf{Trajectory} and the data structure used 
by the third-party library can be made via \emph{yupiwrap}. This approach enables users 
to seamlessly use the resources needed from either library.

\subsubsection{Integration with traja}\label{subsub:integration_with_traja}

\emph{Traja} \textit{Python} package is a toolkit for numerical characterization and 
analysis of moving animal trajectories \citep{shenktraja}. It provides some machine 
learning tools that aren't yet available in \emph{yupi}.

To convert from \emph{yupi} to \emph{traja} let us first consider an arbitrary \emph{yupi} Trajectory:

\begin{verbatim}
    traj = Trajectory(
        x=[0, 1.0, 0.63, -0.37],
        y=[0, 0, 0.98, 1.24])
\end{verbatim} 

\noindent
and then the conversion to a \emph{traja} DataFrame is done by:

\begin{verbatim}
    from yupiwrap import yupi2traja
    traj_traja = yupi2traja(traj)
\end{verbatim} 

Notice that only two-dimensional trajectories can be converted to a \emph{traja} 
object due to its own limitations. The conversion in the opposite direction can 
be done in a similar way by using \textbf{traja2yupi} instead.

\subsubsection{Integration with tracktable}\label{subsub:integration_with_tracktable}

\emph{Tracktable} provides a set of tools for handling two- and 
3-dimensional trajectories even in geospatial coordinate systems 
\citep{tracktable}. The core data structures and algorithms in this 
package are implemented in \textit{C++} for speeding up computation and 
more efficient memory usage.

If we consider the same \emph{yupi} \textbf{Trajectory} from the 
previous example, we can convert it into a \emph{Tracktable} object 
using:

\begin{verbatim}
    from yupiwrap import yupi2tracktable
    traj_track = yupi2tracktable(traj)
\end{verbatim}

In this case, all trajectories with a number of dimensions within 1 to 3 can be converted into \emph{Tracktable} objects. The conversion in the opposite direction can also be done by importing 
\textbf{tracktable2yupi}.

\section{Examples}\label{sec:examples}

In this section, we illustrate the usage of \emph{yupi} through 
different examples that require a complex integration of different modules.
The examples were chosen to showcase the potential of the library
to solve problems that heavily rely on trajectory analysis and its 
extraction from video sources. Most of the examples reproduce core 
results from published research. Others include original approaches to 
verify known properties of different phenomena. All in all, the collection
of examples is designed to provide a quick starting point for new
research projects involving trajectory data, with a particular focus in
environmental modelling.

For simplicity, we omit some technical details related to its implementation. 
However, we provide a detailed version of these examples in the software 
documentation\footnote{Documentation available at
\url{https://yupi.readthedocs.io/en/latest/}} with required multimedia resources 
and source code, available in a repository conceived for \emph{yupi} examples\footnote{Examples available at \url{https://github.com/yupidevs/yupi_examples}}.

\subsection{Identifying environmental properties through tracking and trajectory processing}
\label{sub:ej3}

Visual tracking has proven to be an effective method for the study of 
physical and biological processes. Moreover, indirect measurements 
from the enviroment can also be retrieved from the analysis of the
directly measured trajectories. For instance, tracking techniques have 
empowered researchs on animal behavior, which can improve the effectiveness 
and success of conservation management programs \citep{greggor2019using} 
and gives valuable information about ecosystem changes \citep{rahman2022linking}. 
More especifically, in \citep{yuan2018biological} the authors infered the 
water quality through the analysis of features from fish trajectories. 
Furthermore, the work of \citep{panwar2020aquavision} shows how to automate 
the detection of waste in water bodies using images from the environment. 
In addition, the measurements of river flows have also benefited from the 
tracking of key features using satellital images \citep{gleason2017tracking}.

\begin{figure}[ht]
    \centering
    \includegraphics[width=1\linewidth]{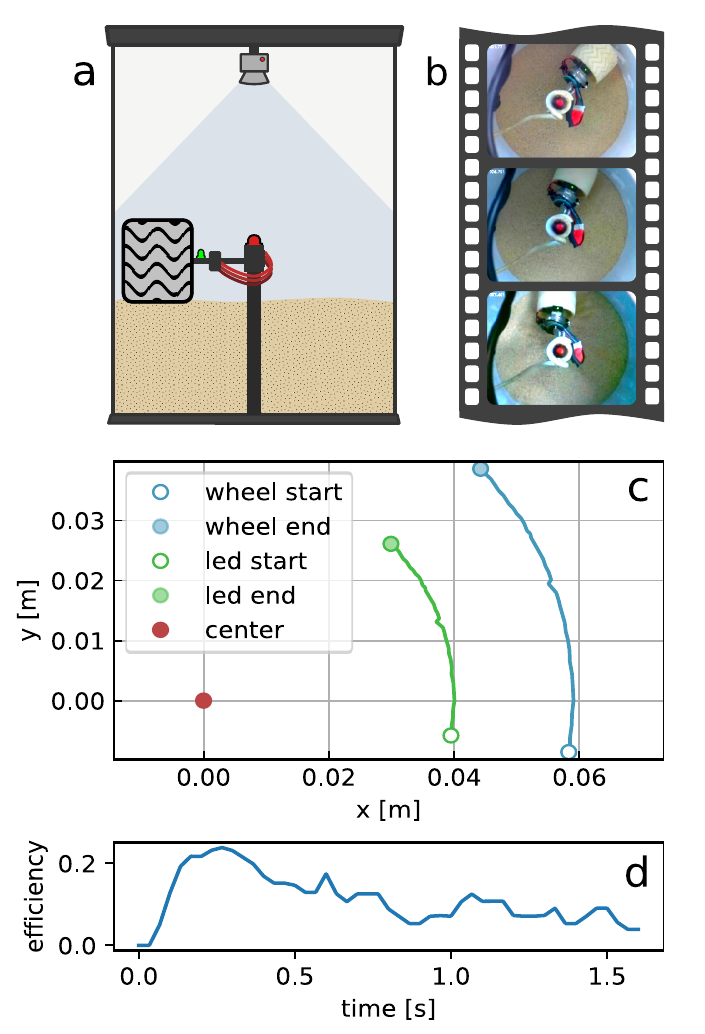}
    \caption{Efficiency of the rolling process computed for a wheel moving across 
    granular material at a constant angular velocity $\omega=4$\,rad/s under a
    gravitational acceleration as the one on Mars. (a) Experimental setup composed
    by the wheel electro-mechanical system to ensure the constant angular velocity 
    and a camera to record its motion. (b) Sample frames from a video where the
    wheel moves around the pivot. (c) Estimated positions of the wheel and the LED
    used as a reference to track it. (d) Estimation of the rolling efficiency
    for a single realization of the experiment, showing a strong consistence
    with the ones reported in the original paper \citep{amigo2019measuring}.}
    \label{fig:C}
\end{figure}

In the work of \citep{stephenson1999vehicle}, it was shown how the stresses of 
turning wheels in the grounds can affect vegetation cover, plant 
health and diversity, as well as reducing underground rhizomes (roots) 
generation. Inspired on these facts, we decided to reproduce
the tracking results from the work of \citep{amigo2019measuring} on the study of 
the motion of vehicles on granular materials. They reported the 
analysis of the trajectories performed by a scaled-size wheel while rolling on 
sand at two different gravitational accelerations, exploiting a frugal instrument
design \citep{viera2017note, altshuler2014settling}. Figure \ref{fig:C}a 
shows a sketch of the instrument where a camera on top captures the 
motion of the wheel while rolling around a pivot. This example was built using 
one of the original videos provided by the authors (see Figure \ref{fig:C}b).

In the video, one observes the wheel forced to move on sand at a fixed angular
velocity. In optimal rolling conditions, one can expect it to move at a
constant linear velocity. However, due to slippage and compaction-decompaction
of the granular soil, the actual linear velocity differs from the one expected
under ideal conditions. To study the factors that affect the wheel motion, the
first step is quantifying how different the rolling process is with respect to
the expected one in ideal conditions. This example focuses on the problem of 
capturing the trajectory of the wheel and computing the efficiency of the 
rolling process.

We start by creating two trackers: one for the central pivot and one for the 
green led attached next to the wheel. Since the central pivot should not move
significantly, we can track it using \textbf{TemplateMatching} algorithm, by
comparing every frame with a template of the object. As the led colors differs
from the rest of the image, we can use \textbf{ColorMatching} algorithm to
track its position.

Both trackers are used among the \textbf{TrackingScenario} to retrieve the 
trajectories of both objects along the video. It is worth to mention that, 
for an accurate estimation, it is required to know the scale factor (i.e.,
the number of pixels required to represent 1\,m). 

By calling the \textbf{track} method, the tracking process should produce
two trajectories (one for each tracker). Notice that these trajectories (i.e.,
${t_\text{led}}$ and ${t_\text{pivot}}$) are referred to a frame of
reference placed on the bottom left corner of the image as shown in Figure \ref{fig:C}.

Then, using the
arithmetic operations from \emph{yupi} it is possible to estimate the 
trajectory of the LED referred to the center pivot by simply subtracting 
them as: $t_\text{led\_centered} = t_\text{led} - t_\text{pivot}$.

Since the LED and the center of the wheel are placed at a constant distance of
0.039\,m, we can estimate the trajectory of the wheel referred to the center 
pivot:

\begin{verbatim}
    wheel_centered = led_centered.copy()
    wheel_centered.add_polar_offset(0.039, 0)
\end{verbatim}

Finally, the trajectory of the wheel referred to its initial position, can be
obtained by subtracting the initial from the final position after completing
the whole trajectory.

\begin{verbatim}
wheel = wheel_centered - wheel_centered.r[0]
\end{verbatim}

Now, assuming no slippage, we can compute the linear velocity as: 
$v_\text{max} = \mathbf{\omega}R$ = $(4 \,\mathrm{rad/s}) \times (0.07 \,\mathrm{m})$ 
and measure the actual linear velocity 
using the trajectory estimated by the tracking process:

\begin{verbatim}
    v_actual = wheel.v.norm
\end{verbatim}

By dividing $v_\text{actual}$ by $v_\text{max}$, we can estimate the 
efficiency of the rolling as described in 
\citep{amigo2019measuring}. The temporal evolution of the efficiency 
for the single experiment can be observed in Figure~\ref{fig:C}d.

We can notice how the linear velocity of the wheel is not constant despite the
constant angular velocity, due to slippage in the terrain. Even when we are
observing only one realization of the experiment, and assuming the angular
velocity of the wheel being perfectly constant, we notice the consistency
of this result with the one reported in the original paper.

Despite the specific nature of this example, it is easy to make a 
straightforward extension of its usage across many other
problems that may require the identification of objects from video sources
and the application of arithmetic operations over trajectories to indirectly
measure any derived quantities. In that regard, we included in the software 
documentation additional examples related to trajectory tracking that partially 
reproduce key results from published research: The work of 
\citep{diaz2020rolling} where the authors study the penetration of objects into 
granular beds; The work of \citep{frayle2017chasing}, where the authors studied 
the capabilities of different image processing algorithms that can be used for 
tracking of the motion of insects under controlled environments and the work of 
\citep{serrano2019autonomous} that extends on the previous one by proposing the design
of a robot able to track millimetric-size walkers in much larger distances by 
tracking both insect and camera simultaneusly.

\subsection{Equation-based simulations: A molecule immerse in a fluid}
\label{sub:ejlysozyme}

Several systems can be explained using stochastic models as the ones shown
in Section \ref{sub:generating}. To accurately describe them, it is 
required to adjust the parameters of the model according to measureable data. 
Next, we will illustrate how to use \emph{yupi} to generate simulated trajectories 
of a lysozyme in water using the Langevin model presented in Section \ref{subsub:langevin}. 
We corroborate that the model correctly predicts the order of magnitude of 
molecule' speed values.

Proteins contribute greatly in environmental processes and are fundamental in 
soil and ecosystem health \citep{li2020goethite}. Since interactions of proteins 
with charged surfaces are important in many applications such as biocompatible 
medical implants \citep{subrahmanyam2002application}, the dynamics of lysozyme and 
its hydration water has been characterized under electric field effects in 
different water environments \citep{favi2014dynamics}. Alongside, the thermal 
velocity for a sizeable particle immerse in water such as a lysozyme molecule at 
room temperature has been estimated to be around $10 \,\mathrm{m/s}$ 
\citep{berg2018random}.

The right hand side of the Langevin equation (\ref{eq:langevin-eq}) can be
interpreted as the net force acting on a particle. This force can be written as
a sum of a viscous force proportional to the particle's velocity (i.e., Stokes'
law with drag parameter, $\gamma=1/\tau$, with $\tau$ a correlation time), and
a noise term, $\sigma\,\boldsymbol{\xi}(t)$, representing the effect of
collisions with the molecules of the fluid. Therefore, (\ref{eq:langevin-eq})
can be written in a slightly different way by noting that there is a relation between the
strength of the fluctuating force, $\sigma$, and the magnitude, $1/\tau$, of
the friction or dissipation, which is known as the Fluctuation-dissipation
theorem \citep{kubo1966fluctuation,srokowski2001stochastic}. Consequently, in terms of
experimental measured quantities and in differential form, the Langevin
equation can be reformulated in the light of stochastic processes by
\begin{subequations}
   \begin{align}
      d\mathbf{v} &= -\frac{1}{\tau} \mathbf{v} dt + \sqrt{\frac{2}{\tau} \left( \frac{kT}{m} \right)} \, d\mathbf{W}
      \label{eq:oh-process-particle-fluid}
      \\
      \frac{1}{\tau} &= \gamma = \frac{\alpha}{m} = \frac{6 \pi \eta a}{m}
      \label{eq:stoke's-law}
   \end{align}
\end{subequations}

\noindent
where $k$ is the Boltzmann constant, $T$ the absolute temperature, and $m$ the
mass of the particle. Equation~\ref{eq:stoke's-law} provides an operational
method to measure the correlation time in terms of the Stoke's coefficient,
$\alpha$, which depends on the radius of the particle, $a$, and the fluid
viscosity, $\eta$.

Lysozyme enzymes are molecules with a high molecular weight ($\sim
10^{4}\,\mathrm{g/mol}$) \citep{colvin1952size}. So, it is reasonable to expect
a brownian behavior in the limit of large time scales when the particle is
subjected to the molecular collisions of the surrounding medium (e.g., an
aqueous medium). Then, Equation~\ref{eq:oh-process-particle-fluid} is a good
choice to use as a model.

\begin{figure}
    \centering
    \includegraphics[width=.9\linewidth]{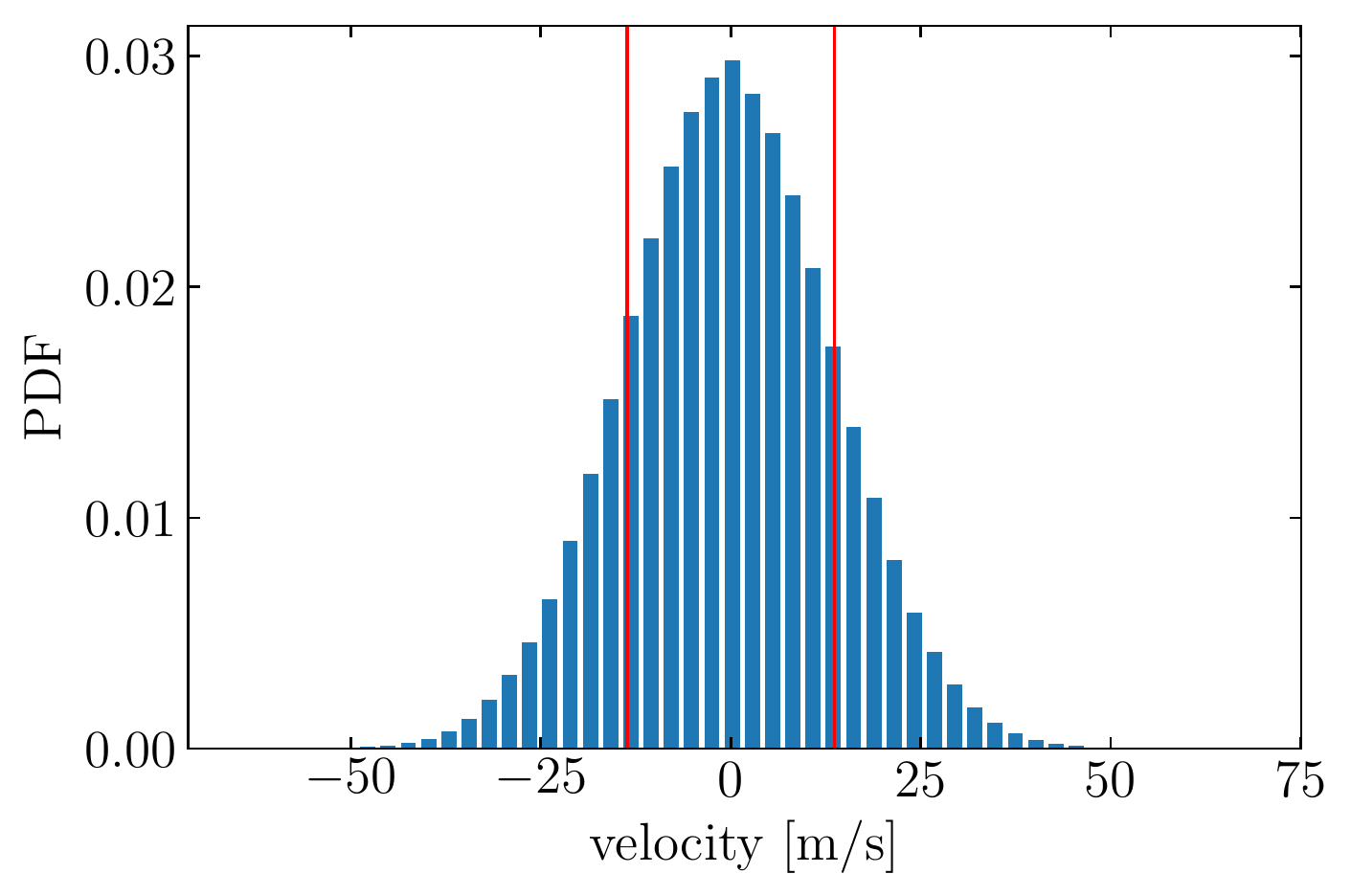}
    \caption{Gaussian velocity distribution predicted by Equation~\ref{eq:oh-process-particle-fluid} for trajectories of a lysozyme molecule in water. Standard deviation, depicted as vertical red lines, coincides with those values in which $\langle v^2 \rangle^{1/2} \sim 10 \,\mathrm{m/s}$.}
    \label{fig:v_pdf}
\end{figure}

By setting the total simulation time \textbf{T}, the dimension \textbf{dim}, the number \textbf{N} and the time step \textbf{dt} of the simulated trajectories, as well as the coefficients \textbf{gamma} and \textbf{sigma} of Equation~\ref{eq:langevin-eq}, we can instantiate the \textbf{LangevinGenerator} class and generate an ensemble of trajectories:
\begin{verbatim}
from yupi.generators import LangevinGenerator 
lg = LangevinGenerator(
    T, dim, N, dt, gamma, sigma)
trajs = lg.generate()
\end{verbatim}

Figure~\ref{fig:v_pdf} shows the velocity probability density function that the 
model predicts. Apart from the typical Gaussian shape that arises when massive 
particles are jiggling, the standard deviation (vertical red lines) is in 
agreement with the previous estimation for the local thermal velocity.

Another equation-based simulation is presented as part of the complementary 
examples provided in \emph{yupi} documentation. It covers the computation of the 
probability density function for displacements at different time instants for 
the case of a one-dimensional process that follows the equations of a Diffusing 
Diffusivity model (see Section \ref{eq:diffdiff}). The example reproduces
important results from the paper presented by \citep{chechkin2017brownian}.

\subsection{Time series analysis: water consumption examination}
\label{sub:water_usage}

This example showcases the usage of \emph{yupi} beyond ``real'' trajectories. 
We reproduce an example from \citep{hipel1994time} in the context of hydrological 
studies by simply treating a time series as an abstract trajectory.

\begin{figure}
    \includegraphics[width=1\linewidth]{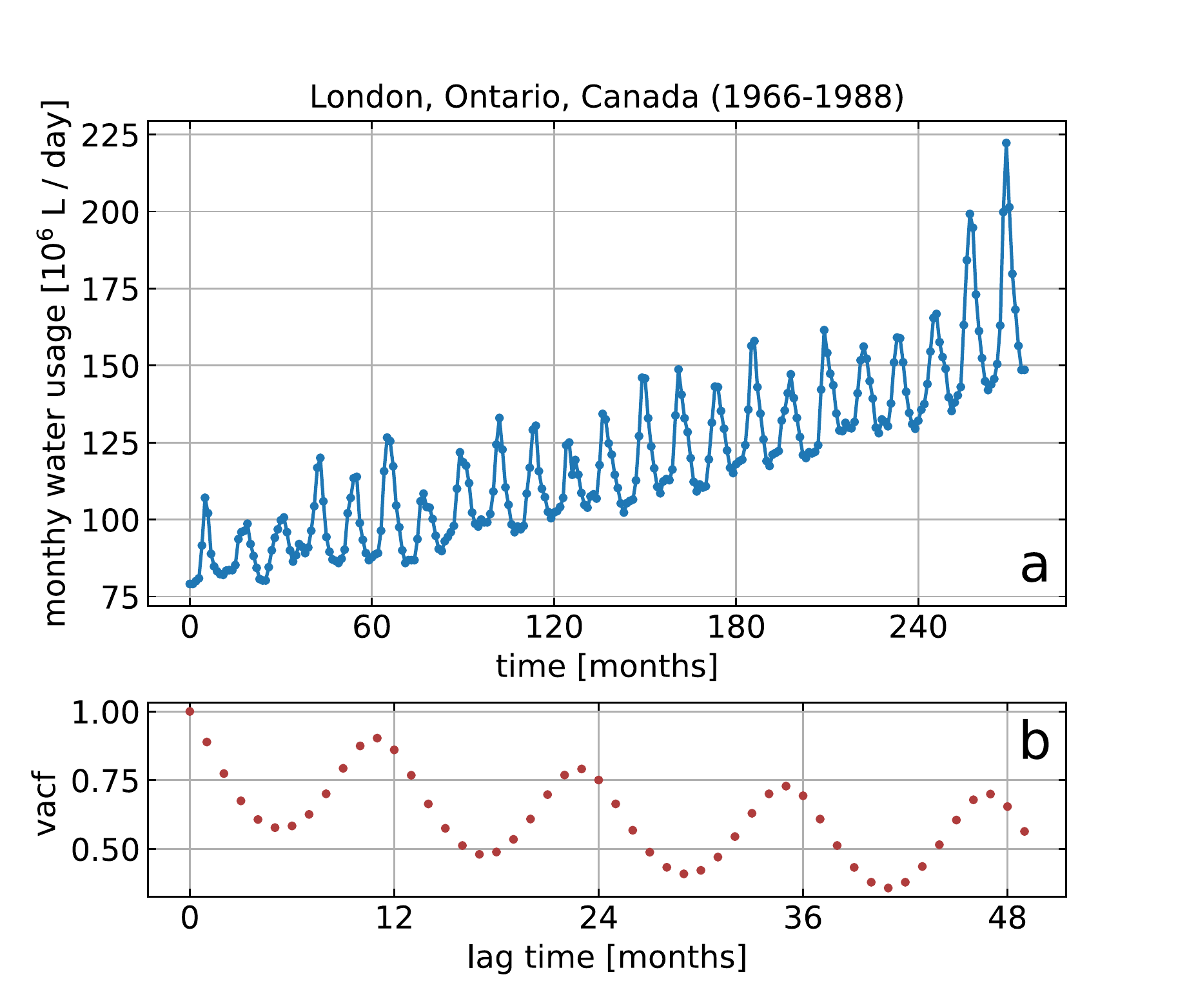}
    \caption{Analysis of water consumption in for the city of London, Ontario, Canada.
    (a) Seasonality of the average monthly water usage (1966-1988). The 
    sinusoidal pattern on top of the linear trend accounts for 
    nonstationarity. 
    (b) Slow decreasing of the normalized autocorrelation 
    function as a function of the lag time. Oscillations are portrayed with 
    a period of a year.
    (Original plots can be found in Figures VI.2/12.4.1 pages 417/440, in 
    \citep{hipel1994time}.)}
    \label{fig:water_usage}
\end{figure}

Seasonal autoregressive integrated moving average 
\\(SARIMA) models\footnote{SARIMA is a forecasting model that supports seasonal (S: seasonal) 
components of a time series that is assumed to depend on its past values (AR: 
autoregressive), past noises (MA: moving average) and resulted from many 
integrations (I: integrated) of some stationary process.} 
are useful for modelling seasonal time series in which the mean and other 
statistics for a given season are not stationary across the years. Some types of 
hydrological time series which are studied in water resources engineering could 
be nonstationary. For example, socio-economic factors such as an increasing of 
population growth in the city of London, Ontario, Canada since the Second World 
War to 1991, caused a greater water demand in the period \citep{hipel1994time}.   
Figure \ref{fig:water_usage}a shows the average monthly water consumption (in 
millions of liters per day) from 1966 to 1988 for this city. The increasing 
trend around which the seasonal data fluctuates reveals nonstationary 
characteristics. As a consequence, autocorrelation analysis was used by 
\citep{hipel1994time} in the design and study of a SARIMA model for the water 
usage time series.

First, let \textbf{waterusage} be the variable in which the time series has been 
stored. Visualization of the data shown in Figure \ref{fig:water_usage}a can be 
done using:
\begin{verbatim}
from yupi import Trajectory
traj = Trajectory(x=np.cumsum(water_usage))
plt.plot(traj.v.x)
\end{verbatim}

Computation and visualization of the autocorrelation function depicted in Figure 
\ref{fig:water_usage}b (see Section \ref{subsub:VACF} for theoretical details) 
can be simply coded as:
\begin{verbatim}
from yupi.stats import vacf
from yupi.graphics import plot_vacf
acf, _ = vacf([traj], time_avg=True, lag=50)
plot_vacf(acf / acf[0], traj.dt, lag=50,
    x_units='months', y_units=None)
\end{verbatim}

\section{Conclusions}\label{sec:conclusion}

This contribution presents \emph{yupi}, a general purpose library for handling 
trajectory data. Our library proposes an integration of tools from different 
fields conceived as a complete solution for research applications related to 
obtaining, processing and analyzing trajectory data. Resources are organized 
in modules according to their nature. However, consistency is guaranteed using 
standardized trajectory data structures across every module.

We have shown the effectiveness of the tool by reproducing results 
reported in a number of research papers. 
We believe the examples illustrating the simplicity of \emph{yupi} 
should enable researchers from different fields to become more proficient in
processing and analyzing trajectories even with minimal programming knowledge.

The current version of \emph{yupi} does not provide 
specific functionalities to process geo-spacial data. Considering the wealth of 
available tools to tackle these specific tasks, we encourage the re-utilization of 
existing approaches for specific use cases by providing an extension to 
simplify two-way conversions of data among some existing trajectory-related 
software libraries.

\section*{Acknowledgements}\label{sec:acknowledgements}

We acknowledge the inspiration received by the coding practices and previous results 
obtained by A. Serrano-Muñoz, which was the main motivation for putting together this 
library as a whole. Also, we would like to thank M. Curbelo for contributions regarding 
the revision of the manuscript.

\printcredits

\bibliographystyle{cas-model2-names}



\end{document}